# Optical read-out of Coulomb staircases in a moiré superlattice via trapped interlayer trions

Hyeonjun Baek,[1,*] Mauro Brotons-Gisbert,[1] Aidan Campbell,[1] Valerio Vitale,[2] Johannes Lischner,[2] K. Watanabe,[3] T. Taniguchi,[4] and B. D. Gerardot[1,*]

[1]*Institute of Photonics and Quantum Sciences, SUPA, Heriot-Watt University, Edinburgh EH14 4AS, UK*
[2]*Departments of Materials and Physics and the Thomas Young Centre for Theory and Simulation of Materials, Imperial College London, South Kensington Campus, London SW7 2AZ, United Kingdom*
[3]*Center for Functional Materials, National Institute for Materials Science, 1-1 Namiki, Tsukuba 305-0044, Japan*
[4]*International Center for Materials Nanoarchitectonics, National Institute for Materials Science, 1-1 Namiki, Tsukuba 305-0044, Japan*

*e-mail address: HJBaek@kist.re.kr, B.D.Gerardot@hw.ac.uk*

Moiré patterns with a superlattice potential can be formed by vertically stacking two layered materials with a relative twist or lattice constant mismatch. In transition metal dichalcogenide based systems, the moiré potential landscape can trap interlayer excitons (IXs) at specific atomic registries. Here we report that spatially isolated trapped IXs in a molybdenum diselenide/tungsten diselenide heterobilayer device provide a sensitive optical probe of carrier filling in their immediate environment. By mapping the spatial positions of individual trapped IX, we are able to spectrally track the emitters as the moiré lattice is filled with excess carriers. Upon initial doping of the heterobilayer, neutral trapped IX form charged IXs (IX trions) uniformly with a binding energy of ~7 meV. Upon further doping, the empty superlattice sites sequentially fill, creating a Coulomb staircase: stepwise changes in the IX trion emission energy due to Coulomb interactions with carriers at nearest neighbour moiré sites. This non-invasive, highly local technique can complement transport and non-local optical sensing techniques to characterise Coulomb interaction energies, visualise charge correlated states, or probe local disorder in a moiré superlattice.

Van der Waals heterostructures can be designed to confine electrons and holes in unique ways. One remarkable approach is to vertically stack two atomically-thin layers of transition metal dichalcogenide (TMD) semiconductors. The relative twist or lattice mismatch between the two layers leads to moiré pattern formation, which modulates the electronic band structure according to atomic registry. Single particle wave-packets can be trapped in the moiré-induced potential pockets with three-fold symmetry, leading to the formation of trapped interlayer excitons (IXs)[1,2]. Trapped IXs, observed so far in MoSe$_2$/WSe$_2$ heterobilayers, have compelling properties. In the limit of low temperature and weak excitation, the confocal photoluminescence (PL) spectra exhibit a few sharp lines (~100 meV linewidths) with strong helical polarization dependent on the atomic registry and the $C_3$ symmetry of the crystal lattice[3,4]. In addition, highly uniform g-factors dependent on the relative layer twist are observed, clear fingerprints of the spin and valley configurations for excitons composed of band-edge electrons and holes.[3-6] The PL emission from an individual trapped IX exhibits photon antibunching due to its quantum nature and, due to its large out-of-plane permanent dipole, is highly tunable in energy with a vertical electric field[7]. With increasing excitation power, the trapped IX density increases and the ensemble emission exhibits broader linewidths ($\geq$ 5 meV) and multiple peaks due to different IX species can arise[8-12], including charged excitons (trions), which can be controlled in gate-tunable devices[9,10,12].

However, it is not yet clear what the relationship of the trapped IXs with the moiré superlattice is. In principle, trapped IXs should form in regular arrays according to the moiré pattern[1,2]. But, because the superlattice periodicity is ~10 nm, visualising emitter positions in the far-field is a significant challenge. Nevertheless, far-field optical spectroscopy has revealed the existence of *intralayer* moiré excitons and hybridization[13,14] and enabled observations of strong electron and hole correlations[15-18], providing evidence that the moiré superlattice can be robust at the micron-scale in TMD heterobilayers. At the nanoscale though, reconstructions, strain, ripples, and other imperfections can affect moiré patterns, as observed with non-destructive local imaging techniques[19,20]. These structural 'imperfections' in a moiré pattern can have considerable impact on the optical properties of IXs[21-23]. But, for trapped IXs that exhibit magneto-optical properties consistent with perfect $C_3$ symmetry at a specific atomic registry, it is unclear how robust the moiré superlattice is and what the local environment of a trapped IX might be.

A trapped exciton can form a highly sensitive electrometer with optical read-out[24,25]. Here we exploit this technique to probe the local charge environment of trapped IX as a function of carrier doping. We show that the trapped IX emission energy is sensitive to the Coulomb interaction

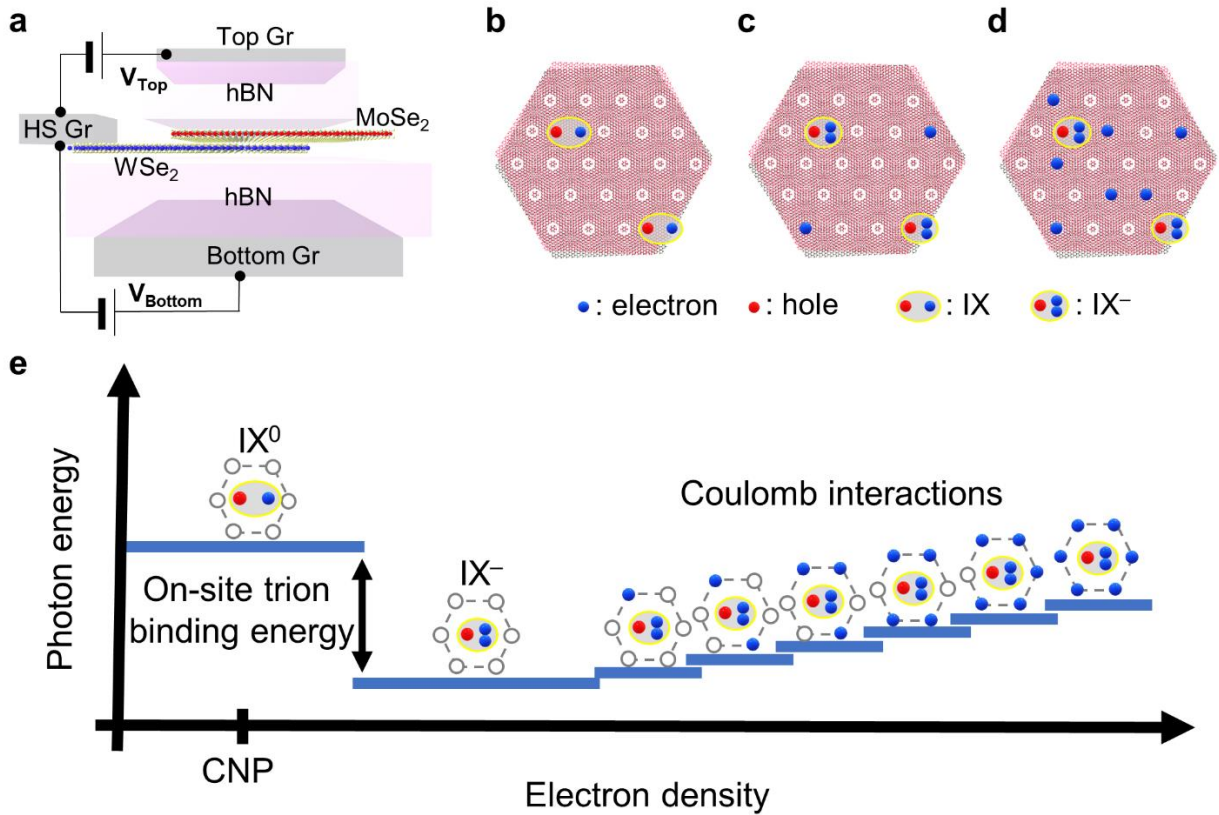

**Figure 1. Schematic illustration of moiré lattice doping. (a)** A schematic of the dual gated device structure for independent control of the electric field and heterobilayer doping. **(b)** A few IX confined in moiré trapping sites in a H-type stacked MoSe$_2$/WSe$_2$ heterobilayer. The white circles represent the trapping sites with atomic configuration of $H_h^h$, where hexagon centres of MoSe$_2$ and WSe$_2$ are aligned. **(c,d)** With increasing doping of the moiré superlattice, IX trions are formed and the number of filled moiré lattice sites increase. **(e)** Illustration of the trapped IX energy versus charge density. As charge density increases, a neutral IX forms an on-site IX trion and the photon emission jumps to lower energy. Successive filling of NN moiré sites leads to a Coulomb staircase. CNP represents the charge neutrality point.

energy of single nearby electrons trapped in the plane of the TMD.

### Electrostatic doping in a moiré superlattice

Our concept is illustrated in Figure 1. Figure 1a shows a sketch of the dual-gated 2H-type stacked MoSe$_2$/WSe$_2$ heterostructure device that allows independent control of doping and electric field. Figures 1b-1c depict a few moiré-trapped IXs at different electron doping densities in the moiré lattice. The white circles represent moiré sites where the hexagon centres of MoSe$_2$ and WSe$_2$ are vertically aligned (i.e. with an $H_h^h$ local atomic registry). At low excitation power, only a few optically generated IXs are formed and confined in the most favourable moiré sites[12], as shown in Fig. 1b. As additional electrons are added, the neutral IXs (IX$^0$) are charged, forming on-site negative trions (IX$^-$), and more empty moiré sites start to fill with excess electrons (Fig. 1c). Upon further electron doping, the number of moiré trapping sites filled by excess electrons increases and the cumulative Coulomb interaction between the trapped IXs and the spatially pinned electrons becomes stronger (Fig. 1d). Figure 1e shows a schematic of the expected change in photon energy for a trapped IX as increasing electron density leads to sequential filling of the six nearest neighbour (NN) moiré sites. At small electron densities, the Fermi level in the device stays within the band gap, and the energy of the localised IX$^0$ remains constant, resulting in an energy plateau around the charge neutrality point (CNP). When the electron density increases and the Fermi energy reaches the bottom of the conduction band, excess electrons start to fill the moiré lattice. The first electrons are expected to occupy the same moiré sites as the trapped IXs due to the positive on-site trion binding energy.[9,10,12] This leads to the formation of an on-site IX$^-$, whose energy is red-shifted compared to the trapped exciton. Upon further electron doping, the increasing density of excess electrons leads to a sequential filling of the NN sites. The repulsive Coulomb interaction between the $IX^-$ and the electrons in the NN sites induces a blue-shift of the $IX^-$ energy. In contrast to conventional semiconductors, where the doped carriers can move freely, the spatial pinning of the additional electrons to the moiré superlattice is expected to result in a discretization of the Coulomb interactions, creating an effective Coulomb blockade effect. Thus, the discrete Coulomb interactions between the localised $IX^-$ and the carriers trapped in neighbouring moiré sites should give rise to a staircase-like blue-shift of the exciton energy, rather than a continuous and smooth shift (see Fig. 1d). We refer to this effect as the Coulomb staircase. These distinguishing spectral features under controlled carrier doping might represent valuable optical probes to read out the charge configuration of the local moiré landscape in TMD heterobilayers.

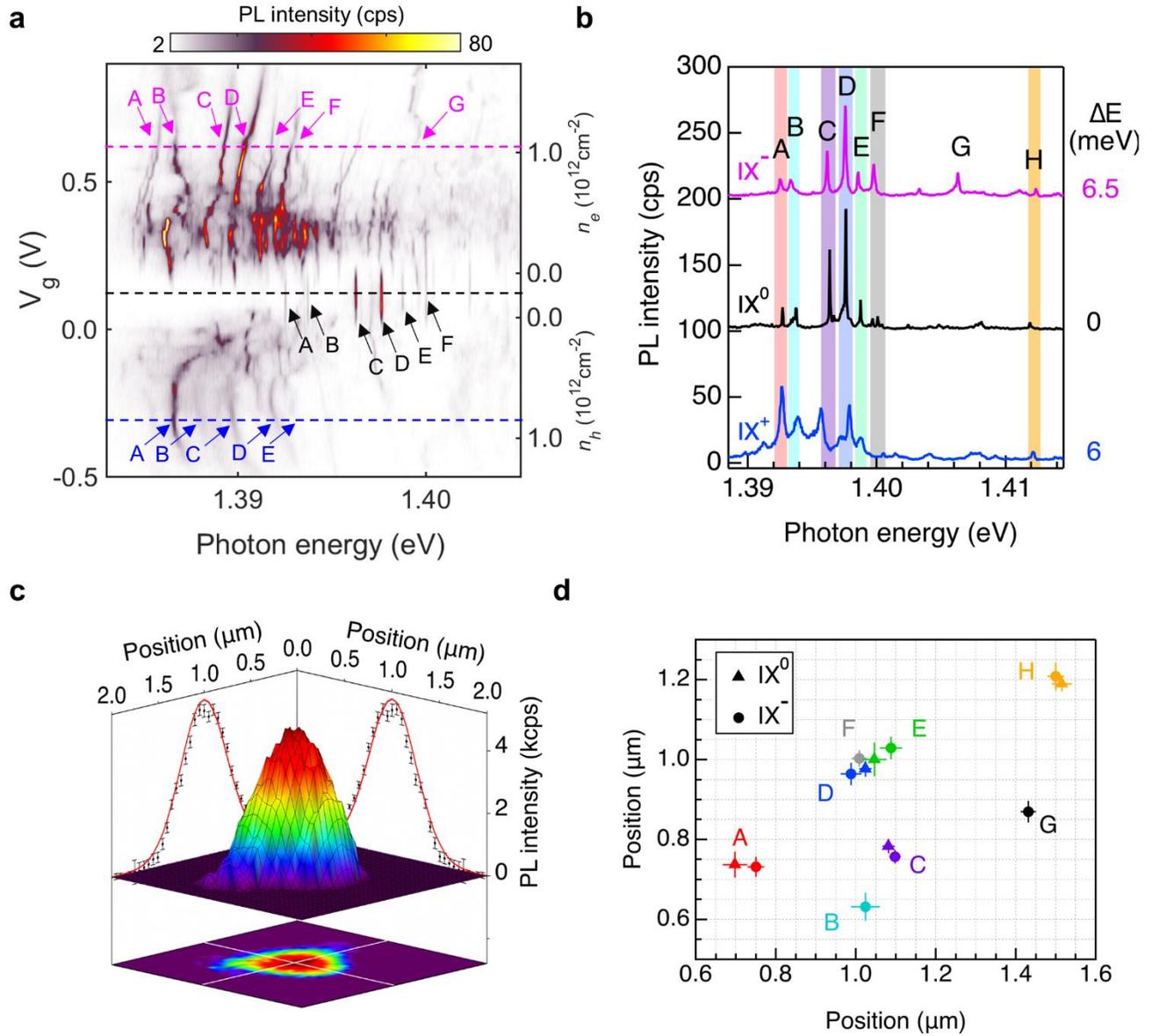

**Figure 2. Doping dependence of moiré IXs. (a)** PL spectrum map as a function of $V_g$. The prominent IX emission peaks are labelled as A-H in each doping region. The dashed lines identify the $V_g$ of the linecuts in (b). **(b)** Representative PL spectra for trapped $IX^0$, $IX^-$, and $IX^+$. For direct comparison with $IX^0$ PL spectrum, the $IX^-$ ($IX^+$) PL spectra are shifted by 6.5 (6) meV in relative photon energy. **(c)** A high-resolution spatial map of the integrated PL intensity of emitter D. From the Gaussian fits, the spatial position can be determined with ~ 25 nm precision. **(d)** Spatial positions of the trapped $IX^0$ (circles) and $IX^-$ (triangles) for various emitters. Error bars are from the standard deviation of repetitive measurements.

We use the same dual-gated 2H-type stacked MoSe$_2$/WSe$_2$ heterostructure device used in previous works[7,12]. hBN layers are used as dielectric spacers, with near identical thickness (~20 nm, see Ref. 7 for detailed information), and graphene is used as electrical contact for the top, bottom, and heterobilayer gates. The carrier density in the device can thus be controlled by applying the same voltage to the top ($V_{Top}$) and bottom ($V_{Bottom}$) gates, minimizing the electric field across the heterobilayer. The 2H stacking configuration was confirmed by the Landé g-factor of the trapped IXs, a clear indicator of the relative valley alignment between the layers hosting the carriers[3-6,12]. Using voltage-dependent reflectance spectroscopy measurements, the quality of the sample and formation of the moiré superlattice is undoubtedly demonstrated: clear signatures of strongly correlated states in both the conduction and valence bands are observed for fractional filling factors of 1/3, 2/3, 1, and 4/3 (see Supplementary Section S1 and S2). In-depth analysis of this experiment is beyond the scope of the current manuscript, but our observations are consistent with recent reports of correlated insulating states in angle aligned WSe$_2$/WS$_2$ heterobilayer samples[15-18]. Based on the applied voltage required to realize specific fractional filling factors, the relative twist angle is determined to be 56.7±0.2° across the sample (seven independent positions are measured). This confirms our estimate of the twist angle from the cleaved edges of MoSe$_2$ and WSe$_2$ (see Ref. 7 and Supplementary Section S3), and our expectation that moiré domain reconstruction is unlikely in our sample[26,27].

**Charged moiré interlayer excitons**

Figure 2a shows a confocal PL spectrum measured at a representative position in the sample (position $P_1$) as a function of the applied gate voltage ($V_g = V_{Top} = V_{Bottom}$). The energy of the CW excitation laser is set to 1.705 eV to resonantly excite the 1s state of intralayer A-excitons of WSe$_2$, and an excitation power of 9 nW is used to ensure only a handful of trapped IXs are optically generated[12]. At $V_g = 0.1$ V, the PL spectrum shows several discrete lines with emission energies in the range 1.39–1.40 eV, in agreement with previously reported values for localised IX$^0$ in 2H-stacked MoSe$_2$/WSe$_2$ heterobilayers[3,4]. As expected for a device with symmetric gates, the emission energy of the trapped excitons remains constant in the neutral region (0.04 < $V_g$ < 0.19 V), confirming a negligible electric field in the direction parallel to the IX permanent electric dipole. We label the peaks in this region A to F. At $V_g > 0.19$ V (i.e. electron doping), the spectrum changes drastically: the overall emission energy red-shifts abruptly by ~7 meV (in agreement with the formation of on-site IX$^-$ trions[12]) and the number of emission peaks increases compared to the neutral charge regime (see Supplementary Section 4). A further increase of $V_g$ results in a combination of a continuous and staircase-like blue-shifts of the emission energies of the localised IX$^-$. At $V_g \approx 0.4$ V the additional PL peaks collapse again into a handful of IX$^-$ (labelled from A to G). At larger biases, the intensity decreases; we do not understand this behaviour but speculate that Auger processes could be involved. The behaviour in the hole-doping regime (negative applied $V_g$), in which we observe the formation of positively-charged IXs (IX$^+$) with an average binding energy of ~6 meV, is generally similar to the n-doped regime. We remark that these spectra as a function of doping are remarkably reproducible: the spectral jumps and slopes observed as the doping is tuned are reproducible numerous times; they are unlikely to be caused by charge noise. Figure 2b shows PL spectra at $V_g = -0.3$ (IX$^+$), 0.1 (IX$^0$) and 0.6 V (IX$^-$) extracted from Fig. 2a (as indicated by the blue, black and pink dashed lines, respectively), where each spectrum has been shifted in energy by the amount ΔE indicated in the right side of the figure. Surprisingly, we find that the number of emission peaks in the spectra of the charged IXs, their relative energies, and their overall relative intensities resemble the PL spectrum observed in the neutral charge regime. We note that this behaviour is not particular to this position in the sample, but is a general behaviour observed at several other positions (see Supplementary Section S5). Such similarities between the PL spectra under negative, neutral, and positive doping conditions suggest that the PL peaks in the different doping regimes originate from the same moiré potential traps, while the spectral jumps as carriers fill the lattice are caused by each emitters' local environment.

To corroborate this hypothesis, we employ a sub-diffraction positioning technique to track the spatial position of emitters A-H as a function of doping. In a confocal microscope, the spatial 2D intensity distribution resulting from a sub-diffraction point-like source (such as a localized IX) is given by the point spread function. The centre of the 2D intensity distribution corresponds to the spatial position of the emitter, which can be located with arbitrarily high precision under the appropriate experimental conditions[28,29]. Our method to determine the spatial positions of the trapped IX involves several steps: first, we carry out 2 x 2 µm$^2$ high-resolution spatial PL maps with a step size of 50 nm centred around the spatial position where the brightest emission is observed. Second, we obtain 2D spatial maps of the integrated PL intensity for each emitter. This is done by fitting the emission peak of each emitter to a Lorentzian function and plotting the resulting intensity as a function of the spatial position on the sample. We note that this process can be carried out efficiently as long as the different emitters can be spectrally resolved and present signal-to-noise ratios > 5. Figure 2c shows an example of the resulting spatial distribution of the integrated PL intensity from a single trapped IX, as obtained for emitter D under n-doped conditions. The position of each trapped exciton and its uncertainty (standard deviation) are then obtained by numerical fitting of the corresponding spatial intensity distribution to a 2D Gaussian[30]. In order to mitigate the effects of possible experimental artifacts in the emitter positioning (such as scanning stage hysteresis or instabilities), the described localization procedure is repeated twice for each $V_g$, and an average position and uncertainty are obtained for each trapped IX. This process allows us to estimate the spatial position of the individual trapped IXs with an average accuracy of 25 nm (i.e. ~20 times smaller than the diameter of our confocal microscopy spot). Figure 2d shows the estimated spatial positions of the emitters shown in Fig. 2b under neutral (triangles) and $n$-doping (circles) conditions. We note that the spectral overlap of peaks B and F at $V_g = 0$ V prevents us from extracting a reliable spatial position for these emitters in the neutral doping regime. The good agreement observed for both the individual and the relative positions of the emitters in the undoped and electron doped regimes corroborates that the spectral lines observed for neutral and $n$-doped conditions originate from the same moiré potential traps. Further, the spatial mapping of the trapped IX positions indicates that, at the exceptionally low density of excitons we optically generate, their position is essentially random: the moiré lattice is not discernible. We speculate that the particular site chosen by the exciton is likely determined by the local environment and perhaps aided by dipolar repulsion effects[31,32].

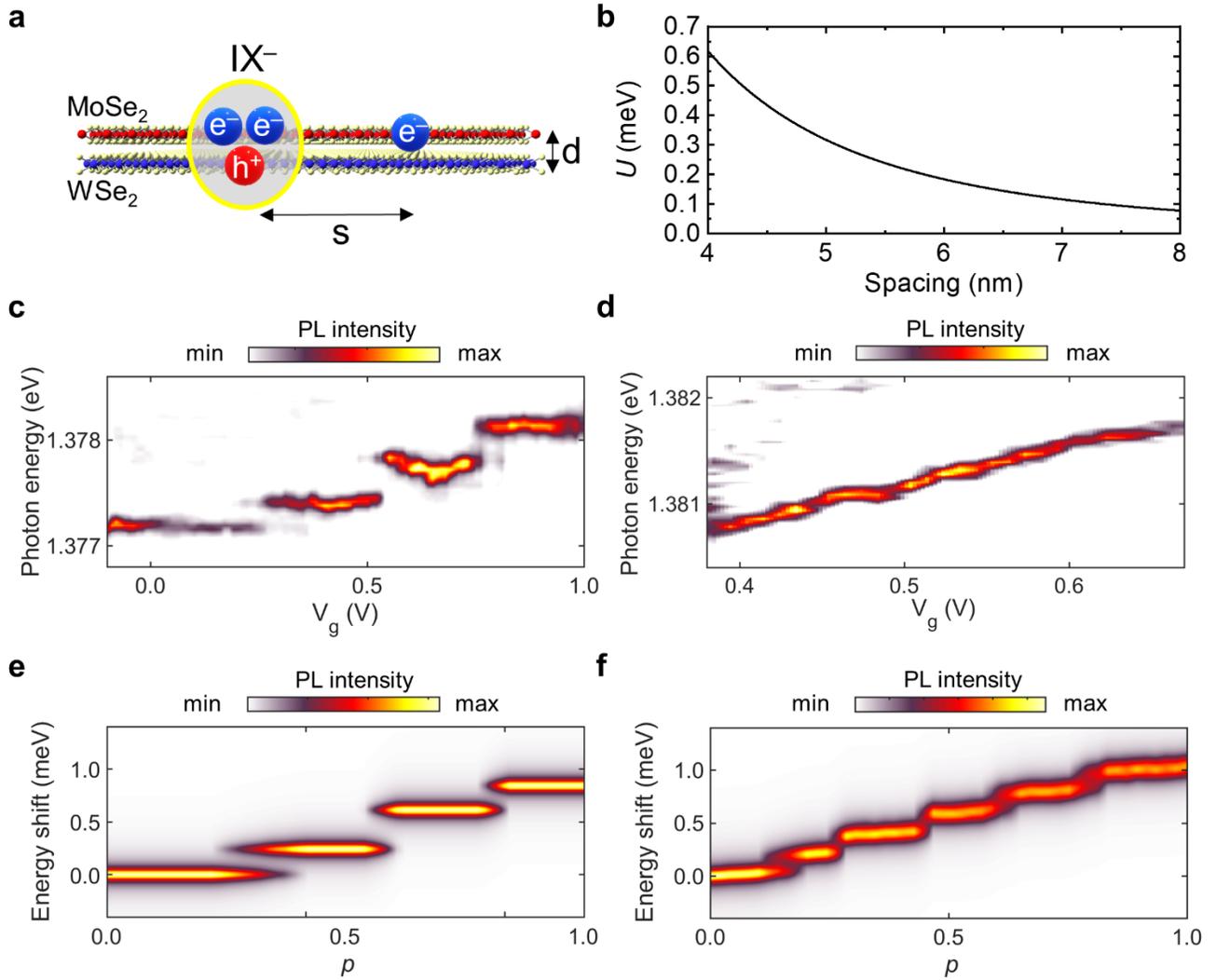

**Figure 3. The Coulomb staircase.** (a) Side-view schematic of a trapped IX⁻ and one electron filled in a NN moiré site. Due to the interlayer distance between the hole and the electron in the IX⁻, the Coulomb interaction between the IX⁻ and a NN electron is repulsive. (b) The calculated Coulomb interaction energy between IX⁻ and a single NN moiré trapped electron as a function of moiré lattice spacing. (c,d) Two representative PL peak shift trends with electron doping. Discrete spectral jumps (c) or a continuous evolution with small jumps (d) are observed. (e,f) Monte Carlo simulation results for discrete (e) and continuous (d) staircase changes. The simulation control parameter, $p$, is swept from 0 to 1 to simulate increasing filling factor. With $p = 0$ (1), all moiré sites are set to be empty (filled).

**Coulomb staircases in a moiré superlattice**

Having determined that the PL emission peaks observed at $V_g = 0.1$ and 0.6 V in Fig. 2a belong to the same moiré-trapped IXs, we now focus on the behaviour of the emission energy of IX$^-$ as a function of electron doping in several spatial positions of the sample. As previously discussed for Fig. 2a, the emission energy of the trapped IX$^-$ blue-shifts with increasing electron doping, featuring a combination of staircase-like and continuous evolution. We speculate that such distinct spectral features have their origin in the Coulomb interaction between the localised IX$^-$ and electrons trapped in neighbouring moiré sites. To investigate this in more detail, we model the Coulomb interaction between a trapped IX$^-$ and a nearby, spatially pinned, excess electron. Figure 3a shows a schematic illustration of the real space configuration of the charges considered in our calculation. The Coulomb interaction ($U$) between the localised IX$^-$ and the nearby electron modifies the energies of both the initial (IX$^-$) and final states (single electron) of the PL emission process. It can therefore be estimated by the equation:

$$U = \frac{e^2}{4\pi\varepsilon_r\varepsilon_0}\left[\frac{1}{s} - \frac{1}{\sqrt{s^2+d^2}}\right], \quad (1)$$

where $e$ is the elementary electron charge, $\varepsilon_0$ is the dielectric permittivity of vacuum, $\varepsilon_r = 4.5$ is the relative dielectric constant of hBN[33,34], $s$ is the moiré period, and $d = 0.5$ nm is the interlayer distance. Moreover, since the thickness of the hBN spacers in our sample (~20 nm) is much larger than the estimated $s$, we ignore the small screening effect caused by the conducting graphene gates. Figure 3b shows the calculated $U$ as a function of $s$. Since $d$ is much smaller than $s$, $U$ decreases as $\sim 1/s^2$.

Figures 3c and 3d present two representative examples of the energy blue-shift observed for IX$^-$ in our heterobilayer as the excess electron density increases (see Supplementary Section S6 for several more examples). As can be seen in Fig. 3c and Supplementary Section S6, increasing $V_g$ leads to a staircase-like blue-shift of the emission energy for some IX$^-$, featuring well-defined discrete spectral jumps. We note that these spectral features are highly reproducible upon electron doping (e.g. multiple scans in $V_g$). The observed energy jumps are in the range of 0.2–0.4 meV, which corresponds to a moiré period of 4.6–5.8 nm and agrees very well with the 5.7 ± 0.4 nm moiré lattice estimated from the twist angle of our heterobilayer (see Supplementary Section S2). These results support the hypothesis that the staircase-like blue-shifts of IX$^-$ originate from the Coulomb interaction of the localised excitons with charge carriers trapped in NN sites.

We underpin the experimental observation of the Coulomb staircases with a Monte Carlo simulation of the IX trion energy as the moiré lattice sites are randomly occupied with electrons (see Methods Section). The model successfully reproduces the prominent features of the experimental data, as exhibited in Fig. 3e and 3f, and can be used to gain further insights into the local environment of each trapped IX. A brief summary of these findings: (i) We find the Coulomb interaction with the NN sites dominates the trapped IX emission energy and accounts for the spectral jumps. (ii) Preferred sequential filling of the NN sites results in reduced overlap of the different energy plateaus. (iii) Inhomogeneities in the spectral jump energy is caused by small (nm-scale) variations in the moiré period for an individual NN site. (iv) The voltage extent of each charging plateau in the energy staircase is given by the occupation probability of the corresponding NN site. (v) Since $U$ decreases as $\sim 1/s^2$, long-range Coulomb interactions from a single electron occupying non-NN sites has a small affect. However, the sum of such interactions from filling many non-NN sites leads to the continuous evolution (as shown in Fig 3d and 3f). We refer to Supplementary Section S7 for full details about the model.

**Polarization of charged moiré interlayer excitons**

To further investigate the properties of the individual trapped IXs, we monitor their degree of circular polarisation (DOCP) as a function of carrier doping. We define the DOCP as $\frac{I_{\sigma^-} - I_{\sigma^+}}{I_{\sigma^-} + I_{\sigma^+}}$, where $I_{\sigma^+}$ ($I_{\sigma^-}$) is the intensity of right (left) circularly-polarised emission. Figure 4a shows the DOCP as a function of $V_g$ obtained for position P$_1$ under $\sigma^-$ polarised laser excitation. Representative PL spectra with $\sigma^+$ (blue) and $\sigma^-$ (red) collection polarisation are presented in Fig. 4b for $n$-doped (top), undoped (middle), and $p$-doped (bottom) conditions. In Fig 4a, one can observe that the doping map has almost no charge noise and is highly reproducible. The exception is one trapped IX$^+$ emitter, which appears as the blue and red peaks at ~1.3866 eV in Fig. 4a due to a slightly different energy in the $\sigma^+$ and $\sigma^-$ maps. We observe the trapped IXs exhibit a strong co-polarisation with the excitation laser for both undoped and electron doping conditions, with an estimated DOCP of 0.7 (0.86) at $V_g = 0.1$ (0.56 V). In stark contrast, the DOCP of the trapped IXs reduces to almost 0 in the hole doped regime. We interpret the drastic change in the DOCP for hole doping as an indication that the photo-excited holes in WSe$_2$ preserve the valley polarisation of the excitation laser, while the valley polarisation of the excitation laser field is lost in the photo-excited electrons. Figures 4c and 4d show schematic configurations of IX$^-$ and IX$^+$ under $\sigma^-$ excitation for an H-type stacked heterobilayer. Due to the type-II band alignment, the excess electron (hole) of IX$^-$ (IX$^+$) resides in the MoSe$_2$ (WSe$_2$) layer. As a consequence of Pauli exclusion, the additional carrier occupies the opposite valley to that containing the initial IX$^0$. Therefore, since electrically-injected electrons (holes) populate both ±K valleys for IX$^-$ (IX$^+$), the DOCP of the trion state is given by the conservation of the valley polarisation of the optically-created hole (electron) in WSe$_2$ (MoSe$_2$). The large DOCP values observed for IX$^-$ in the electron doped regime suggest that the valley polarisation of the excitation laser can be robustly preserved by the excited hole in WSe$_2$. In contrast, the absence of DOCP for hole-doping indicates that the valley polarisation of the excitation laser is totally lost for the photo-excited electrons in MoSe$_2$. Such contrasting behaviour for photo-excited carriers in MoSe$_2$ and WSe$_2$ originates from the different intervalley scattering processes in these materials. As previously reported, the valley relaxation time in monolayer MoSe$_2$ is much shorter than that in WSe$_2$, leading to valley depolarisation[35,36]. Additionally, the slight increase in DOCP from 0.79 for IX$^0$ to 0.86 for IX$^-$ can be also explained by the faster scattering rate in MoSe$_2$. For example, in the case of $\sigma^-$ excitation, the probability of occupancy for the conduction band at the –K valley for IX$^0$ decreases due to the intervalley scattering of electrons in MoSe$_2$ limiting the recombination rate of IX at the –K valley. On the other hand, for IX$^-$, the scattered electrons can be compensated by electrically doped ones, and the recovered recombination rate leads to increased DOCP since it is proportional to $\frac{1}{(1+\tau/\tau_r)}$, where $\tau$ is the exciton decay time, and $\tau_r$ is the valley relaxation time[37]. Moreover, it is worth noting that the overall behaviour of the DOCP is similar under resonant excitation of MoSe$_2$ A-excitons (see Supplementary Section S9 and S10). This suggests that the valley polarisation is preserved exclusively by holes in WSe$_2$ even under resonant excitation to MoSe$_2$.

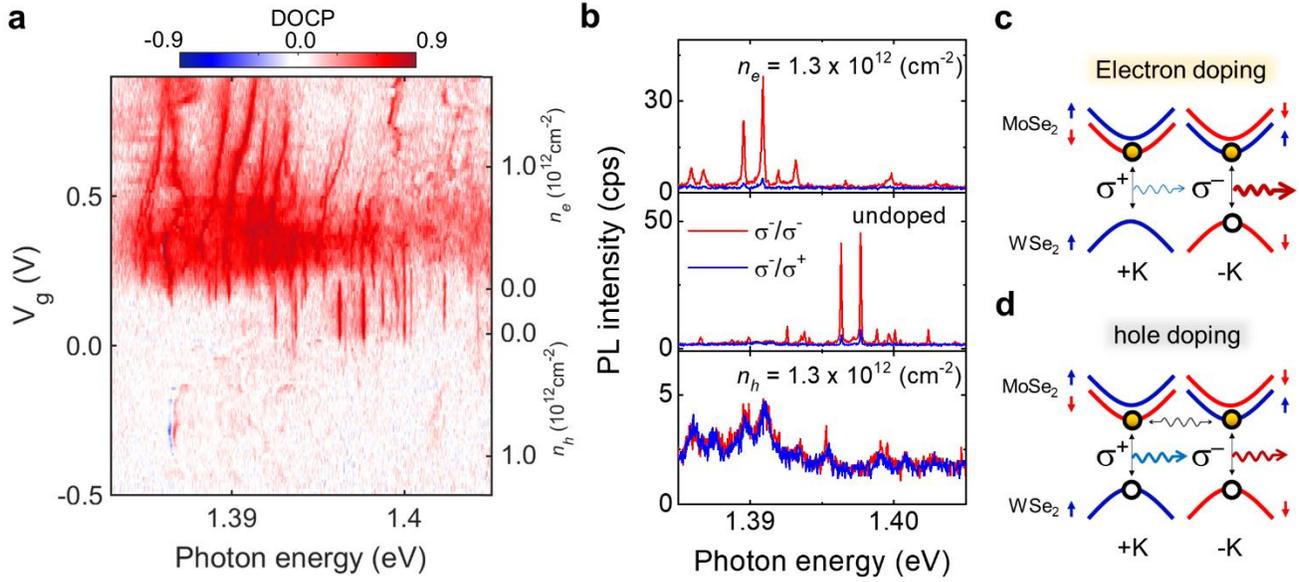

**Figure 4. DOCP as a function of doping.** (a) DOCP map as a function of $V_g$. The map is obtained at the same spatial position for the data in Fig. 2. (b) Representative PL spectra at each doping condition. The red and blue represent co- and cross-polarised PL emission, respectively. A DOCP of nearly 0 can be found with hole doping. (c,d) Energy band diagrams for IX with electron (c) or hole (d) doping conditions with $\sigma^-$ excitation, respectively. For hole doping, intervalley scattering of the electron is represented by a wavy arrow.

**Magneto-optical properties of charged moiré interlayer excitons**

Finally, we investigate the magneto-optical properties of the localised IX under different doping conditions. Figures 5a-c show the PL spectrum of the trapped IXs at position (P$_1$) for applied $V_g$ of 0 (IX$^0$), 0.6 (IX$^-$), and –0.4 V (IX$^+$), respectively, as a function of the applied out-of-plane magnetic field (Faraday geometry). We observe that, similar to the behaviour reported for localised IX$^0$ [3,4,7], both IX$^+$ and IX$^-$ show linear Zeeman splittings with no observable fine-structure. Linear fits of the measured Zeeman splittings reveal Landé g-factors of –16.10, –15.75, and –16.37 for IX$^0$, IX$^-$, and IX$^+$, respectively, as shown in Figs. 5d-f, which agree well with the effective g-factor expected for spin-triplet optical transitions in 2H-stacked MoSe$_2$/WSe$_2$ hetero-bilayers[4,6].

**Conclusions**

In summary, we report here the deterministic creation of a low density of moiré trapped IX trions in a charge tunable 2H-type stacked MoSe$_2$/WSe$_2$ heterostructure device. We spectroscopically investigate their properties and exploit their sensitivity to their immediate environment to

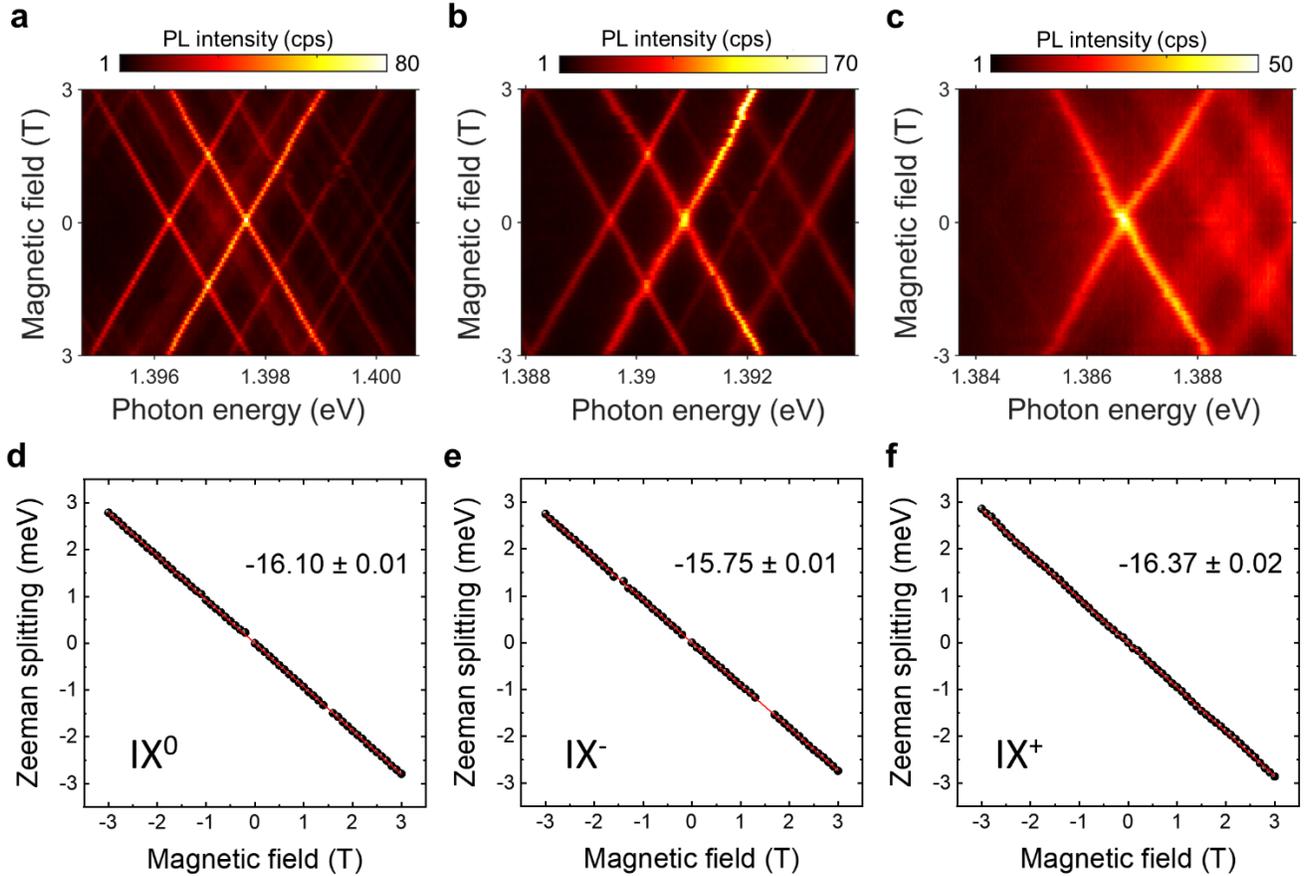

**Figure 5. Magneto-PL characteristics of neutral and charged IX. (a-c)** The PL of $IX^0$, $IX^-$, and $IX^+$ ($V_g = 0$, 0.6, and –0.4 V, respectively) as magnetic field is swept from –3 to 3 T. The excitation laser and PL collection are both linearly polarised. **(d-f)** Plots of the Zeeman splitting versus magnetic field for $IX^0$, $IX^-$, and $IX^+$, yielding g-factors of –16.10, –15.75, and –16.37, respectively.

characterise the doping of the moiré superlattice. First, by spatially tracking their positions, we are able to spectrally track individual emitters as a function of doping. This enables us to unambiguously identify the on-site trion binding energy, which is highly uniform for numerous emitters, and probe the spectral jumps that arise with filling of the moiré superlattice. We observe several examples of the Coulomb staircase: stepwise changes in the IX trion emission energy due to Coulomb interactions with carriers at nearest neighbour moiré sites. We perform a Monte Carlo simulation to better understand the effects of short-range moiré site uniformity and long-range interactions. These results demonstrate a non-invasive, highly local technique to characterise the moiré superlattice. An interesting prospect would be to combine this approach with complementary techniques to probe the local disorder, homogeneity, and reconstruction in a moiré lattice. For example, the two types of trapped IX trion peak shifts in Fig. 3c and d show either well-defined discrete spectral jumps or a more continuous energy shift upon doping, respectively. These two general responses suggest that the local lattice in the vicinity of the trapped IXs has different properties, likely caused by disorder. For the IX peak in Fig. 3c, the small effect of long-range Coulomb interactions from a single electron occupying non-NN sites is negligible, which implies that the moiré site responsible of the IX trapping and its nearby sites have much higher probability of charge occupation compared to sites farther away. On the other hand, moiré sites in the region of the emitter in Fig. 3d can be considered quite homogeneous, allowing a random charge filling of both NN sites and non-NN sites with equal probability. Combining this optical read-out technique with other surface probe microscopy techniques that provide structural information to correlate distinct features of local disorder (e.g. strain, ripples, reconstruction, and other imperfections) in a moiré superlattice [21,38,39] would provide valuable new insights into moiré systems. Furthermore, an exciting prospect would be to locally probe long-range charge-ordered states in a fractionally filled moiré lattice, such as correlated Mott insulators and Wigner crystals[15-17]. The optical-readout of

Coulomb interactions between a single moiré-trapped IX and its neighbouring charges presents a viable route to visualise the formation and Coulomb interaction energies of such charge-ordered states in TMD heterobilayers, although the signatures can be complex (See Supplementary Section 8).

**Methods**
**PL spectroscopy**

A confocal microscope with an objective lens with a numerical aperture of 0.82 is used for PL measurements. The device is loaded into a closed-cycle cryostat (4 K) with a superconducting magnet. A continuous wave Ti:sapphire laser is used to resonantly excite $MoSe_2$ and $WSe_2$ A excitons at $\lambda$ = 759.6 and 727.0 nm, respectively. The PL emission is dispersed in a 500-mm focal length spectrometer and detected by nitrogen-cooled charge-coupled device with a spectral resolution of ~70 μeV at $\lambda$ = 900 nm for 1200 lines/mm. Polarization-resolved PL is measured by changing the relative angle between a quarter–wave plate and a linear polarizer both for laser excitation PL collection.

**Monte Carlo simulation for Coulomb staircases**

We simulate the change of emission energy of a stationary trapped exciton as a function of carrier density using a Monte Carlo model. Our model shares many features with that of Houel *et. al*[25]. We comment that we do not simulate the Hubbard model here, e.g. we only account for $U$, not for the hopping probability $t$, and that our model does not take into account interactions between the excess electrons. The occupation of an electron at each trapping site $i$ is determined randomly and the resulting $U_i$ caused by the occupied lattice sites is calculated. A control parameter ($p$), swept from 0 to 1, is introduced to simulate the doping density. This value is multiplied with a weighting factor $w_i$, which varies from moiré site to site and accounts for the probability of the different sites to trap the injected excess carriers due to inhomogeneity among moiré sites. Then $w_i p$ is compared to a random number ($r$) in the range $0 \leq r \leq 1$. If $w_i p \geq r$, the moiré site is considered to be filled by an electron, otherwise the site is set to be empty. Therefore, all sites are filled when $p$ reaches 1. This process is conducted at each trapping site, and $r$ is newly regenerated at every comparison. In this way, the occupation state of the neighbouring moiré trapping sites is established, and the total Coulomb energy $U_{tot}$ from the occupied sites is calculated as

$U_{tot} = \sum_i U_i$, with $U_i = \frac{e^2}{4\pi\varepsilon_r\varepsilon_0} \left[ \frac{1}{s_i} - \frac{1}{\sqrt{s_i^2+d^2}} \right]$, as discussed

in equation (1). A PL spectrum is then created assuming a Lorentzian line shape with a width of 100 μeV and a central energy $U_{tot}$. This process is repeated $N$ (typically 100–1000) times at a given $p$ resulting in $N$ PL spectra, which are summed together to obtain a final PL spectrum. Finally, by changing $p$ from 0 to 1 we are able to simulate PL spectra at different doping levels.

## ACKNOWLEDGEMENTS

This work is supported by the EPSRC (grant no. EP/P029892/1 and EP/L015110/1), the ERC (grant no. 725920) and the EU Horizon 2020 research and innovation program under grant agreement no. 820423. Growth of hBN crystals by K.W. and T.T. was supported by the Elemental Strategy Initiative conducted by the MEXT, Japan, Grant No. JPMXP0112101001, JSPSKAKENHI Grant No. JP20H00354, and the CREST(JPMJCR15F3), JST. B.D.G. is supported by a Wolf-son Merit Award from the Royal Society and a Chair in Emerging Technology from the Royal Academy of Engineering.